\documentstyle[preprint,aps,epsf]{revtex}

\newcommand{\dnn}[1]{_{\rm \scriptscriptstyle #1}}
\newcommand{\ket}[1]{| \, #1 \rangle}
\newcommand{\bra}[1]{ \langle #1 \,  |}
\newcommand{\proj}[1]{\ket{#1}\bra{#1}}

\newcommand{\pproj}[2]{\ket{#1}\bra{#1}\otimes\ket{#2}\bra{#2} }

\newcommand{\PT}[2]{#2^{\rm T_{#1}}}
\newcommand{\beq}{\begin{equation}}
\newcommand{\eeq}{\end{equation}}
\newcommand{\enscar}[1]{{\cal L_E}(#1)}
\newcommand{\rank}[1]{{\cal R}(#1)}
\newtheorem{lem}{Lemma}
\newtheorem{theo}{Theorem}

\title{Optimal Decompositions of Barely Separable States}
\author{David P. DiVincenzo$^1$,
Barbara M. Terhal$^2$, and Ashish V. Thapliyal$^3$}
\address{\vspace*{1.2ex}
        \hspace*{0.5ex}{$^1$IBM Research Division, Yorktown Heights, NY 10598, USA, {\tt divince@watson.ibm.com}}}
\address{\vspace*{1.2ex}
        \hspace*{0.5ex}{$^2$ ITF, Universiteit van Amsterdam,Valckenierstraat 65, 1018 XE Amsterdam, and \\
CWI, Kruislaan 413, 1098 SJ Amsterdam, The Netherlands, {\tt terhal@wins.uva.nl}}}
\address{\vspace*{1.2ex}
        \hspace*{0.5ex}{$^3$ Dept. of Physics, Univ. of
California, Santa Barbara, CA 93106, USA, {\tt ash@physics.ucsb.edu}}}

\date{\today}
\begin{document}
\maketitle

\begin{abstract}
Two families of bipartite mixed quantum states are studied for which
it is proved that the number of members in the optimal-decomposition
ensemble --- the ensemble realizing the entanglement of formation ---
is greater than the rank of the mixed state.  We find examples for
which the number of states in this optimal ensemble can be larger than
the rank by an arbitrarily large factor.  In one case the proof relies
on the fact that the partial transpose of the mixed state has zero
eigenvalues; in the other case the result arises from the properties
of product bases that are completable only by embedding in a larger
Hilbert space.
\end{abstract}


\pagebreak

\section{Introduction}

The work of recent years has given us an extensive understanding
of the entanglement of pure bipartite quantum states.
While there are
still many open questions about the entanglement of finite collections
of quantum states \cite{JP}, a rather complete understanding of
`asymptotic' entanglement, that of a large number of copies of a
quantum state, has emerged:
a pure state
is unentangled if and only if the state can be written in a product
form $\ket{\Psi}=\ket{\Psi_A}\ket{\Psi_B}$.  The single good
quantitative measure of entanglement is
$E=S(\rho_A)=S({\rm Tr}_B\proj{\Psi})$, where $S$ is the von Neumann
entropy.  And, a collection of bipartite pure states with total
entanglement $E$ can be reversibly interconverted into any other
collection of pure states with entanglement $E$ by purely local
operations.

However, despite much recent effort, we cannot claim to have such a
complete understanding of quantum entanglement for bipartite mixed
states.  Its characterization has a remarkably greater complexity and
richness than the pure-state case: There is {\em not}, except in very
simple cases, an unambiguous way to say if a mixed state is entangled
or not.  There is {\em no} single good quantitative measure of
mixed-state entanglement.  And, it seems that entanglement is
irreversibly lost when one attempts to convert it from one mixed-state
embodiment to another.

Much of this difficulty can be traced to the basic fact
\cite{Sch,Houst} that there is no single way of viewing a mixed
quantum state as an ensemble of pure states.  In fact, we know that
there are infinitely many such representations, and we have previously
noted that in general these pure-state ensembles exhibit entirely
different entanglement properties.  For example, the completely mixed
state of two qubits is equally well described as an ensemble of
product basis states (no entanglement) or as an ensemble of the four
Bell states (all maximally entangled).

It seems that no single measure of entanglement for mixed states is
correct, but many different ones are useful depending on the
situation.  The ensemble decomposition of a mixed state with the
maximal entanglement is useful in situations where the two parties
holding the state can be given aid by a third party to extract a pure
state with the greatest entanglement; we have termed the average
pure-state entanglement of this ensemble the ``entanglement of
assistance'' \cite{div:fuc:mab:smo:tha:uhl:98}. Another,
operational characterization of entanglement is the ``distillable
entanglement'' $D$ \cite{ben:ber:pop:sch:96,bdsw}, the average number
of maximally entangled singlet pairs that can be extracted from many
copies of the mixed state by local operations and classical
communication.  Yet another way of quantifying entanglement related to
$D$ has been proposed \cite{ved:ple:rip:kni:97} in which the minimum
distance from the set of separable mixed states is taken as the
measure of entanglement.

Finally, the entanglement measure on which we focus in this paper is
the ``entanglement of formation'' \cite{ben:ber:pop:sch:96,bdsw}.  It
is the average pure-state entanglement of the ensemble which has {\em
minimal} entanglement that describes the mixed state.  Thus this is
dual, in an operational sense, to the entanglement of assistance.  It
plays several other roles as well: it is converse, in some sense, to
the distillable entanglement, in that it gives the number of EPR
singlet pairs needed to create the mixed state by local operations.
It, like the measure of entanglement in
Ref. \cite{ved:ple:rip:kni:97}, provides bounds on the distillable
entanglement, and thus on other quantities such as the quantum
capacity of noisy channels that are of great current interest in
quantum information theory.

Thus, we believe that a complete characterization of the mathematical
properties of the entanglement of formation should be valuable in the
continued development of quantum information theory.  In this paper we
give new results on one particular feature of the entanglement of
formation, the least number of states needed to make up a
minimal-entanglement ensemble of a mixed state.  (In \cite{uhl} such
optimal decompositions of mixed quantum states have already been
considered, but with respect to a function related to, but different
from, the entanglement of formation.)  Determining such optimal
decompositions gives information about the minimal-complexity
procedures for creating a mixed state from a supply of EPR singlets.
But besides the operational significance of our results, we believe
that the characterizations we provide are of greater significance on
account of the light they shed on the complexity and richness of the
mathematical structure of this important concept in quantum
information theory.

\section{Entanglement Basics}

Let $\rho$ be a density matrix on the bipartite Hilbert space ${\cal H}_A
\otimes {\cal H}_B$ and let ${\cal E}_{\rho}=\{p_i,\ket{\psi_i}\}$
with $p_i>0$ be an ensemble into which $\rho$ can be decomposed:
\beq
\rho=\sum_i p_i \ket{\psi_i} \bra{\psi_i}.
\eeq
The entanglement of formation \cite{bdsw} of $\rho$ is defined as \cite{foot1}
\beq
E(\rho)=\min_{{\cal E}_{\rho}}\sum_i p_i E(\ket{\psi_i}\bra{\psi_i}),
\label{eform}
\eeq
where
\beq
E(\ket{\psi}\bra{\psi})=S({\rm Tr}_A \ket{\psi}\bra{\psi})=S({\rm Tr}_B \ket{\psi}\bra{\psi}),
\eeq
where $S(.)$ is the von Neumann entropy:
\beq
S(\rho)=-\mbox{Tr}\,\rho \log \rho.
\eeq

The minimization in Eq. (\ref{eform}) makes an analytical computation
of the entanglement of formation of mixed states a nontrivial task.
Only in a bipartite Hilbert space $2 \otimes 2$ has
the problem of determining the entanglement of formation of any
density matrix been completely solved, in the work of Wootters
\cite{woot}.  Uhlmann \cite{uhl} has shown that every bipartite
density matrix $\rho$ admits an optimal decomposition, that is, the
one that achieves the entanglement of formation $E$ of $\rho$
(Eq. (\ref{eform})), with at least $\rank{\rho}$ and at most
$\rank{\rho}^2$ different pure states, where $\rank{\rho}$ is the rank
of the density matrix.  We call the number of different pure states in an
ensemble that forms a decomposition of a density matrix $\rho$ the
{\em cardinality} of the ensemble.  We say that the {\em optimal ensemble
cardinality} of a separable state $\rho$, which we denote by
$\enscar{\rho}$, is $k$ if at least $k$ different pure states are
required for a separable decomposition of $\rho$.  Since the number of
states must at least be sufficient to span the support of $\rho$,
\beq
\enscar{\rho} \geq \rank{\rho},
\label{excee}
\eeq
directly giving Uhlmann's lower bound.  However, it has been a open
question whether there are states for which the optimal decomposition
has more than $\rank{\rho}$ different pure states.  Note that the
states in the decomposition of a density matrix $\rho$ are always in
the range of $\rho$.  This means that if $\enscar{\rho} > \rank{\rho}$
the states in the optimal decomposition will be linearly dependent.

In this paper we present two sets of examples of separable bipartite
states $\rho$ for which we prove that the cardinality of the optimal
decomposition of $\rho$ exceeds $\rank{\rho}$. These are the first
examples of such states.  Both types of examples can be found in
principle in arbitrary high dimensions.


It is useful to classify states according to their behavior under
partial transposition. Let $\PT{B}{\rho}=({\bf 1}_A \otimes T) \rho$ where 
$T$ is transposition of a matrix in a chosen basis. $\rho$ is 
{\em positive under partial transposition} (PPT) if $\PT{B}{\rho}$ is a 
density matrix, i.e., it has no negative eigenvalues. If $\rho$ is {\em negative under partial
transposition} (NPT) then $\PT{B}{\rho}$ has at least one negative
eigenvalue.  It is known that for $2 \otimes 2$ and $ 2 \otimes 3$
systems, PPT is a necessary and sufficient condition for separability
\cite{horo1}. For a bipartite state in a Hilbert space of arbitrary
dimension, PPT is a necessary condition for separability
\cite{per:96}.

\section{Separable States At The Boundary}

In this section we show that if a separable state and its partial
transpose have unequal ranks, then one of them must have its optimal
ensemble cardinality greater than its rank. From this result we prove that
partial transposes of full-rank separable states that lie on the
boundary of the set of PPT states have optimal ensemble cardinality greater
than their ranks.  Finally we give examples of such states for any $n
\otimes n$ system.

We start with the following straightforward observation:
\begin{lem}
Let $\rho$ be a separable state on ${\cal H}_A \otimes {\cal
H}_B$. Let $\PT{B}{\rho}=({\bf 1}_A \otimes T) \rho$ where $T$ is
transposition in a chosen basis. Then
\beq
\enscar{\rho}=\enscar{\PT{B}{\rho}}.
\eeq
\label{lemequal}
\end{lem}

{\em Proof\ } We prove that $\enscar{\PT{B}{\rho}} \leq \enscar{\rho}$
and $\enscar{\PT{B}{\rho}} \geq \enscar{\rho}$. Since $\rho$ is
separable, its optimal decomposition involves only product states:
\beq
\rho=\sum_{i=1}^{\enscar{\rho}} p_i \pproj{\psi_i}{\phi_i}.
\label{dec1}
\eeq
Then it follows that
\beq
\PT{B}{\rho}=\sum_{i=1}^{\enscar{\rho}} p_i \pproj{\psi_i}{\phi_i^*},
\label{dec2}
\eeq
and thus $\enscar{\PT{B}{\rho}}$ is at most $\enscar{\rho}$. By
performing the partial transpose again on $\PT{B}{\rho}$ we can prove
the inequality in the other direction.
$\Box$

We have seen that the optimal ensemble cardinality is invariant under
partial transposition. The rank of a density matrix $\rho$ is not
necessarily invariant under partial transposition. We can draw the
following conclusion:

\begin{theo}
Let $\rho$ be a separable state on ${\cal H}_A \otimes {\cal H}_B$.  If
\beq
\rank{\rho} \neq \rank{\PT{B}{\rho}} \enspace,
\label{note}
\eeq
then either $\rho$ has the property that $\enscar{\rho} > \rank{\rho}$ or
$\PT{B}{\rho}$ has the property that $\enscar{\PT{B}{\rho}} > \rank{\PT{B}{\rho}}$.
\label{theorank}
\end{theo}

{\em Proof\ } This follows directly from Lemma \ref{lemequal} and
Eq. (\ref{excee}).
$\Box$

\begin{figure}[htbp]
\epsfxsize=12cm
\epsfbox{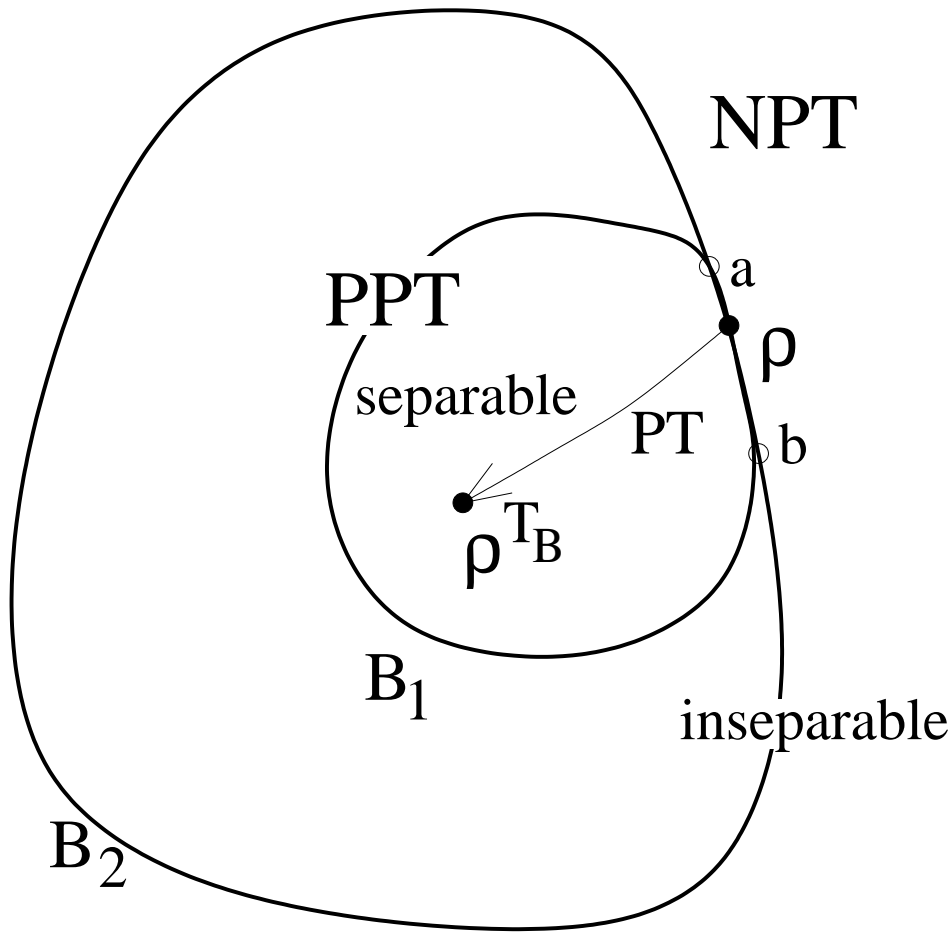}
\caption{Illustration of the construction of Lemma
\protect\ref{boundfull}.  The PPT, NPT, separable, and inseparable
sets of density matrices are indicated. $B_1$ is the boundary between
separable and inseparable, $B_2$ between PPT and NPT.  For $2\otimes
2$ and $2\otimes 3$, $B_1=B_2$, but in general they do not coincide.
The conditions of the Lemma are met when $\rho$ has full rank, is
separable, and is on the boundary $B_2$ so that its partial transpose
$\PT{B}{\rho}$ has some zero eigenvalues.  $\rho$ must therefore be at
a place where $B_1$ and $B_2$ coincide, such as the arc $ab$.
$\PT{B}{\rho}$ is separable; also, it cannot be on the boundary $B_2$,
because its partial transpose, which is $\rho$, has strictly positive
eigenvalues.  However, it is possible that $\PT{B}{\rho}$ could sit on
the boundary $B_1$ where $B_1$ and $B_2$ do not coincide.}
\label{fig1} 
\end{figure}

Where do we find separable states $\rho$ with the property
Eq. (\ref{note})?  For this we look at full-rank separable states that
lie on the boundary of the set of PPT states. The following lemma,
illustrated by Fig. \ref{fig1}, looks into this:

\begin{lem}
Let $\rho$ be a separable state on ${\cal H}_A \otimes {\cal H}_B$ with
full rank, $\rank{\rho}=\dim {\cal H}_A \otimes {\cal H}_B$. If 
$\rho$ lies on the boundary of the set of PPT states, then
\beq
\rank{\rho} > \rank{\PT{B}{\rho}}.
\eeq
\label{boundfull}
\end{lem}

{\em Proof\ } 
The set of PPT states ${\cal S\dnn{PPT}}=\{\rho \mid \PT{B}{\rho} \ge 0 \}$
is a closed convex set. For the separable states $\rho$ on the boundary of 
this set, the state $\PT{B}{\rho}$ has at least one eigenvalue which is zero, 
as $\rho$ is arbitary close to entangled states $\rho_E$ for which 
$\PT{B}{\rho_E}$ has at least one {\em negative} eigenvalue. 
Thus $\PT{B}{\rho}$ does not have full rank, and for full-rank states $\rho$ this implies $\rank{\rho} > \rank{\PT{B}{\rho}}$. $\Box$

%

One can remark the following: In $2 \otimes 2$ and $2 \otimes 3$ all
entangled density matrices have the property that $\PT{B}{\rho}
\not\geq 0$ \cite{horo1}. Therefore any separable density matrix
$\rho$ that is on the boundary of the set of separable states and has
full rank, fulfills the conditions of Lemma \ref{boundfull}. With
Theorem \ref{theorank} it follows that the partial transposes of these
density matrices have the desired property, i.e.,
$\enscar{\PT{B}{\rho}} > \rank{\PT{B}{\rho}}$.

A final comment about the results of this section: Eqs. (\ref{dec1})
and (\ref{dec2}) show that for separable $\rho$, $\rho=\PT{B}{\rho}$
if all the states $\phi_i$ are real in some local basis, so obviously
the ranks of $\rho$ and $\PT{B}{\rho}$ are equal.  Thus, any state
$\rho$ satisfying Lemma \ref{boundfull} must have complex state
vectors in any separable decomposition in any local basis.  Note that
even real density matrices $\rho$ sometimes have optimal
decompositions which require complex vectors \cite{bdsw1}.  Readers
may find it surprising that even for complex vectors, there exist sets
$\ket{\psi_i}\ket{\phi_i}$ for which the set
$\ket{\psi_i}\ket{\phi_i^*}$ spans a space of a different dimension;
but this is exactly the consequence of Lemma \ref{boundfull}.

\subsection{Examples}
\label{ex1}

The generalized Werner state \cite{horodistill} in $n \otimes n$ is defined 
as
\begin{equation}
\rho\dnn{W}(f)={f}\proj{\Psi^+}+\frac{1-f}{n^2} {\bf 1}_{n^2} \enspace ,
\label{genwern}
\end{equation}
where $\ket{\Psi^+}=\frac{1}{\sqrt{n}}\sum_{i=0}^{n-1} \ket{ii}$ is
the maximally entangled state in $n \otimes n$ and $0 \leq f \leq 1$. 
Let
\beq
\rho(f)=\PT{B}{\rho\dnn{W}(f)}$, on$\  {\cal H}_n\otimes{\cal H}_n.\ \
\eeq
It has 
been shown by Horodecki and Horodecki \cite{horodistill} that the state
$\rho\dnn{W}(f)$ is separable for $0\le f\le 1/n$. On the other hand,
for $1/n<f\le 1$, $\rho(f)$ is not positive semidefinite. Therefore the 
state $\rho\dnn{W}(1/n)$ lies at the PPT-NPT boundary. Upon inspection of 
Eq. (\ref{genwern}) we see that the rank of $\rho\dnn{W}(1/n)$ is full, 
$\rank{\rho\dnn{W}(1/n)}=n^2$. Thus we can use Theorem \ref{theorank} and 
Lemma \ref{boundfull}. 

It is easy to show by direct calculation that $\rho(1/n)$ has exactly
$n(n-1)/2$ zero eigenvalues and hence its rank is $n(n+1)/2$.
Therefore by Theorem \ref{theorank} and Lemma \ref{boundfull}, we find
that $\rho(1/n)$ has an optimal ensemble cardinality of at least $n^2$
whereas the $\rank{\rho(1/n)}=n(n+1)/2$.  For this state, the ratio
$\frac{\enscar{\rho(1/n)}} {\rank{\rho(1/n)}}$ can be as large as 2
when $n$ tends to $\infty$.  This ratio could be made higher if there
exists a $\rho$ in $n \otimes n$ for which $\PT{B}{\rho}$ can have
more than $n(n-1)/2$ eigenvalues; we have no indication that this is
possible.  But for the state $\rho(1/n)^{\otimes k}$, $k$ tensor
copies of the above state, the ratio of ${\cal L}_{{\cal E}}$ to
${\cal R}$ can be made arbitrarily large (going to $2^k$ for
$n\rightarrow\infty$).

\section{Barely Completable Sets of Product States}

In \cite{upb1,upb2} the notions of an unextendible product
basis and an uncompletable product basis were introduced.  A product
basis is a set of $k$ separable orthogonal pure states in $n\otimes
m$.  Considering the case $k<nm$, the basis is unextendible if there
are no additional pure product states orthogonal to all the members of
the basis; it is uncompletable (in $n\otimes m$) if the number of such
additional states is less than $nm-k$. In \cite{upb1,upb2} it is shown
that the completely mixed state $\rho$ on the Hibert space
complementary to the space spanned by the unextendible product basis
is entangled, but has the property that ${\rho^{\rm T_B}}$ is positive
semidefinite.  In the following, we will need the notion of a local
extension of a bipartite Hilbert space ${\cal H}={\cal H}_A \otimes
{\cal H}_B$: a local extension of ${\cal H}$ is a Hilbert space ${\cal
H}'=({\cal H}_A \oplus {\cal H}'_A) \otimes ({\cal H}_B \oplus {\cal
H}'_B)$.

In \cite{upb1} and \cite{upb2} the notions of an unextendible product
basis and an uncompletable product basis were introduced
\cite{expl}. It was shown how to construct, from an unextendible
product basis, a bipartite entangled state $\rho$ for which
$\PT{B}{\rho}$ is positive semidefinite.  We will need the notion of a
local extension of a bipartite Hilbert space ${\cal H}={\cal H}_A
\otimes {\cal H}_B$: A local extension of ${\cal H}$ is a Hilbert
space ${\cal H}'=({\cal H}_A \oplus {\cal H}'_A) \otimes ({\cal H}_B
\oplus {\cal H}'_B)$.

\begin{theo}
Let $\{\ket{\alpha_i} \otimes \ket{\beta_i}\}_{i=1}^{|S|}$ be a partial
product basis $S$ in ${\cal H}={\cal H}_A \otimes {\cal H}_B$.
If $S$ is uncompletable in ${\cal H}$, but $S$ is completable in some
local extension ${\cal H}'$ of ${\cal H}$, then $\rho_S$ defined as
\beq
\rho_S=\frac{1}{\dim {\cal H}-|S|}\left({\bf 1}-\sum_{i=1}^{|S|}
\pproj{\alpha_i}{\beta_i}\right),
\eeq
has the property that
\beq
\enscar{\rho_S} > \rank{\rho_S}.
\eeq
\label{uncomprank}
\end{theo}

{\em Proof\ } As the set of states $S$ is completable in a local
extension of ${\cal H}$, the state $\rho_S$ is separable, by Lemma 2
of \cite{upb1}.  The idea of this Lemma 2 is that the completion of the
set $\{\ket{\alpha_i} \otimes \ket{\beta_i}\}_{i=1}^{|S|}$ in ${\cal
H}'$ give rise to a separable state
\beq
\rho_S'=\frac{1}{\dim {\cal H'}-|S|}\left({\bf 1'}-\sum_{i=1}^{|S|}
\pproj{\alpha_i}{\beta_i}\right).
\eeq
But $\rho_S$ is obtained from $\rho_S'$ by local projections on to
${\cal H}_A$ and ${\cal H}_B$ and therefore $\rho_S$ is separable
as well. However, since $\rho_S$ is uncompletable in ${\cal
H}$, $\rho_S$ cannot be represented as an ensemble of
orthogonal product states of cardinality $\rank{\rho_S}$. Thus any optimal
decomposition of $\rho_S$ must use non-orthogonal product states. The
von Neumann entropy of $\rho_S$ is equal to $S(\rho_S)=\log
\rank{\rho_S}$ as $\rho_S$ is the identity on a space of dimension
$\rank{\rho_S}$. In order to achieve this entropy, the optimal
decomposition of $\rho_S$ has to use more than $\rank{\rho_S}$ product
states, or
\beq
\enscar{\rho_S} > \rank{\rho_S},
\eeq
because any density matrix $\rho$ which is a mixture of only $n$
non-orthogonal states has entropy strictly less than $\log n$
bits. $\Box$

\subsection{An Example}

In \cite{upb1} an example was given of a set of orthogonal product
states on $3 \otimes 4$ that is not completable in $3 \otimes 4$, but
is completable in $3 \otimes 5$.  We reproduce the states here:
Consider the states $\vec{v}_i\otimes \vec{w}_i,\;\; i=0,\ldots,4$
with $\vec{v}_i$ defined as
\beq
\vec{v}_i=N (\cos{{ 2 \pi i \over 5}},\sin{{2 \pi i}\over 5},h),\;\;
i=0,\ldots,4,
\label{defP}
\eeq
with $h={1 \over 2} \sqrt{1+\sqrt{5}}$ and
$N=2/\sqrt{5+\sqrt{5}}$. The states $\vec{w}_j$ are defined as
\begin{eqnarray}
\nonumber\vec{w}_j&=N'(\sqrt{\cos(\pi/5)} \cos(2j\pi/5),\sqrt{\cos(\pi/5)} \sin(2j\pi/5),\\
&\sqrt{\cos(2\pi/5)}\cos(4j\pi/5),\sqrt{\cos(2\pi/5)}\sin(4j\pi/5)),
\end{eqnarray}
with normalization $N'=\sqrt{2/\sqrt{5}}$. Note that $\vec{w}_j^T
\vec{w}_{j+1}=0$ (addition mod $5$). One can show that this set,
albeit extendible on $3 \otimes 4$, is not completable: One can at
most add three vectors like $\vec{v}_0 \otimes
(\vec{w}_0,\vec{w}_1,\vec{w}_4)^{\perp}$, $\vec{v}_3 \otimes
(\vec{w}_2,\vec{w}_3,\vec{w}_4)^{\perp}$ and
$(\vec{v}_0,\vec{v}_3)^{\perp} \otimes
(\vec{w}_1,\vec{w}_2,\vec{w}_4)^{\perp}$.  The completion of this set
in $3 \otimes 5$ is particularly simple, being given by the following
ten states:
\beq
\begin{array}{lr}
(\vec{v}_1,\vec{v}_4)^{\perp} \otimes \vec{x}_0, & \vec{v}_0 \otimes (\vec{w}_0^{\perp} \in \mbox{span}(\vec{x}_4,\vec{x}_1)), \\
(\vec{v}_0,\vec{v}_2)^{\perp} \otimes \vec{x}_1, & \vec{v}_1 \otimes (\vec{w}_1^{\perp} \in \mbox{span}(\vec{x}_0,\vec{x}_2)), \\
(\vec{v}_1,\vec{v}_3)^{\perp} \otimes \vec{x}_2, & \vec{v}_2 \otimes (\vec{w}_2^{\perp} \in \mbox{span}(\vec{x}_1,\vec{x}_3)), \\
(\vec{v}_2,\vec{v}_4)^{\perp} \otimes \vec{x}_3, & \vec{v}_3 \otimes (\vec{w}_3^{\perp} \in \mbox{span}(\vec{x}_2,\vec{x}_4)), \\
(\vec{v}_0,\vec{v}_3)^{\perp} \otimes \vec{x}_4, & \vec{v}_4 \otimes (\vec{w}_4^{\perp} \in \mbox{span}(\vec{x}_3,\vec{x}_0)).
\end{array}
\label{complPO}
\eeq
The state $\rho_S$ on $3  \otimes  4$ has rank seven, but the
separable decomposition consists of ten non-orthogonal states obtained
by projecting the orthogonal states of the completion,
Eq. (\ref{complPO}), back into the $3  \otimes  4$ Hilbert space. It
is not known whether there exists a separable decomposition with more
than seven but with fewer than ten states.

\section{Discussion}

The results presented here on the minimum cardinality of optimal ensembles
raises a large number of tantalizing questions; we would like to know
this cardinality for all possible mixed states.  So far, our rigorous
results apply only to separable states.  There is some empirical
evidence that for entangled mixed states as well (in fact, for states
arising in the theory of unextendible product bases), the optimal
ensembles can have a cardinality greater than the rank \cite{JSpc}.  But we
have found no techniques for proving any results for inseparable mixed
states.  We would also like to know whether there are cases for which
the Uhlmann upper bound of ${\cal R}(\rho)^2$ is attained.  The states
shown above are still far from this; for the states at the end of
Sec. \ref{ex1} with $n=2$ and any $k$, ${\cal L}_{{\cal E}}={\cal
R}^{{\log 4}\over{\log 3}}$.  Finally we note that all the rigorous
results we have pertain to cases where ${\cal L}_{{\cal E}}$, while
greater than ${\cal R}$, never exceeds the Hilbert space dimension.
Is there some reason that ${\cal L}_{{\cal E}} $ can never exceed this
dimension?
In conclusion, we have shown two different families of unentangled
mixed states for which the minimal number of pure states in an optimal
minimal-entanglement ensemble is provably greater than the rank of the
mixed state.  In both cases the proofs are possible because the mixed
state is marginally separable, in the first case because the partial
transpose of the state has zero eigenvalues, and in the second because
the state is defined as the complement of a barely completable product
basis.

\section*{acknowledgments}

We are grateful to Peter Shor for the construction of the barely
uncompletable product-basis example used here.  We thank Charles
Bennett, Richard Jozsa, John Smolin, and Armin Uhlmann 
for helpful discussions.  AVT
would like to thank David D. Awschalom for his invaluable support
without which it would have been impossible to work in this exciting
field.  This work has been supported in part by the Army Research
Office under contract numbers DAAG55-98-C-0041 and
DAAG55-98-1-0366. AVT and BMT would like to thank IBM Research for
logistical support during their visits to the IBM Thomas J. Watson
Research Center.

\begin{references}
\bibitem{foot1} It has been pointed out that a more satisfactory
definition of the entanglement of formation would be
\begin{equation}
E(\rho)=\min_{N,{\cal E}_{\rho^{\otimes N}}}{1\over N}
\sum_i p_i E(\ket{\psi_i}\bra{\psi_i}).
\label{eform+}
\end{equation}
Eqs. (\protect\ref{eform+}) and (\protect\ref{eform}) are equal if the
entanglement of formation is additive, but this property has not been
proved.  The analysis of this paper concerns the original definition
of entanglement of formation, Eq. (\protect\ref{eform}).
\bibitem{bdsw1} Eq.(30) of \protect\cite{bdsw} gives a decomposition
with eight members for any mixture of Bell states. However it can be
easily seen that appropriate sets of four of these give optimal-cardinality
ensembles with at least one complex vector.
\bibitem{expl} A product basis is a set of $k$ separable orthogonal
pure states in $n\otimes m$.  Considering the case $k<nm$, the basis
is unextendible if these are no additional pure product states orthogonal
to all the members of the basis; it is uncompletable (in $n\otimes m$) if 
the number of such additional states is less than $nm-k$.
\end{references}
\end{document}